\documentclass[a4paper]{article}
\usepackage{graphicx}
\usepackage{overpic}
\usepackage{verbatim} 

\usepackage{setspace}

\newenvironment{centreFigure}
{
\begin{center}
\hspace*{-0.5in}\hspace*{-\marginparwidth}\begin{minipage}[b]{0.9\paperwidth}
\begin{center}
}
{
\end{center}
\end{minipage}
\end{center}
}

\begin{document}

\title{Fourier Transforms as a tool for Analysis of Hadron-Hadron Collisions.}

\author{M.~Campanelli and J.~W.~Monk\\\
\\\
 Dept. of Physics \& Astronomy\\\
 University College London\\\
 Gower Street, London, England}

\maketitle

\begin{abstract}
Hadronic final states in hadron-hadron collisions are often studied by clustering final state hadrons into jets, each jet approximately corresponding to a hard parton.  The typical jet size in a high energy hadron collision is between $0.4$ and $1.0$ in $\eta-\phi$.  On the other hand, there may be structures of interest in an event that are of a different scale to the jet size.  For example, to a first approximation the underlying event is a uniform emission of radiation spanning the entire detector, colour connection effects between hard partons may fill the region between a jet and the proton remnant and hadronisation effects may extend beyond the jets.  We consider the possibility of performing a Fourier decomposition on individual events in order to produce a power spectrum of the transverse energy radiated at different angular scales.  We attempt to identify correlations in the emission of radiation over distances ranging from the full detector size to approximately 0.2 in $\eta-\phi$.  As a demonstration of this technique we apply it  to a comparison of di-jet events produced with and without a colour connection between the jets.
\end{abstract}

\section{Introduction}

The study of collective event-wide distributions, with the aim of extracting QCD properties, has been carried out over several generations of lepton and hadron colliders \cite{Fox:1978vu}. We believe that within a single event it should be possible to isolate both some global characteristics as well as the smaller scale radiation in order to simultaneously discriminate between various hard scattering and hadronisation models. 

This problem of separating objects of different size is not unique to high energy collider physics.  In the field of cosmology, for example, a key observation is the angular size of correlations in the temperature and polarisation of the cosmic microwave background (CMB).  This is studied  by decomposing the image of the CMB into a set of spherical harmonics in order to produce the well known angular power spectrum plot \cite{Nolta:2008ih}.  Fourier transforms are also used in many other fields of physics and signal processing, but have so far remained largely ignored in the analysis of high energy collider physics data.  In this paper we propose the idea of performing a Fourier decomposition on hadron collider events with the goal of separating via their different Fourier coefficients the large scale features such as the underlying event from smaller features such as hadronisation, showering and the hard jets.  

As a test-case, and because it was the event topology that inspired us to consider using a Fourier transform, we will examine some Monte Carlo samples of di-jets produced diffractively at a collision energy of 10~TeV.  The di-jets are produced through the exchange of a colour singlet object.  Colour singlet interactions occur in both vector boson exchange or, with a much larger cross section, hard diffraction with forward jets.   Di-jets produced via the exchange of a colour singlet have a feature that there is a region in pseudo-rapidity, $\eta$, between the jets with suppressed emission of radiation.  The low-activity region between the jets is commonly known as a rapidity gap.  This jet-gap-jet topology is present in both diffractive and vector boson fusion events and could be used to suppress background events produced through the exchange of coloured QCD objects.  

\begin{figure}
\begin{centreFigure}


\begin{minipage}[c]{0.25\linewidth}

\includegraphics[width=\linewidth]{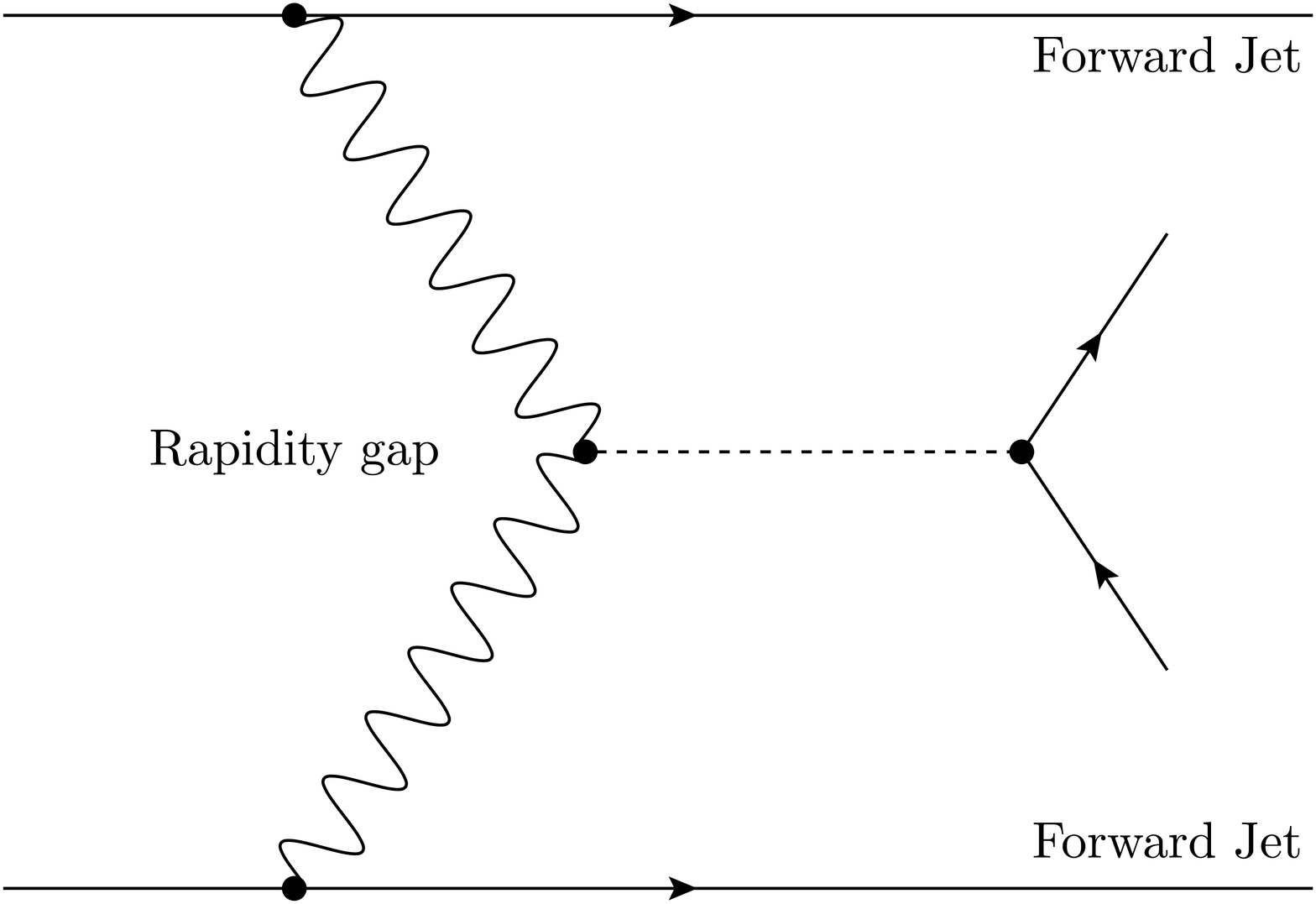}
\end{minipage}
\hspace{0.5in}
\begin{minipage}[c]{0.25\linewidth}
\includegraphics[width = \linewidth]{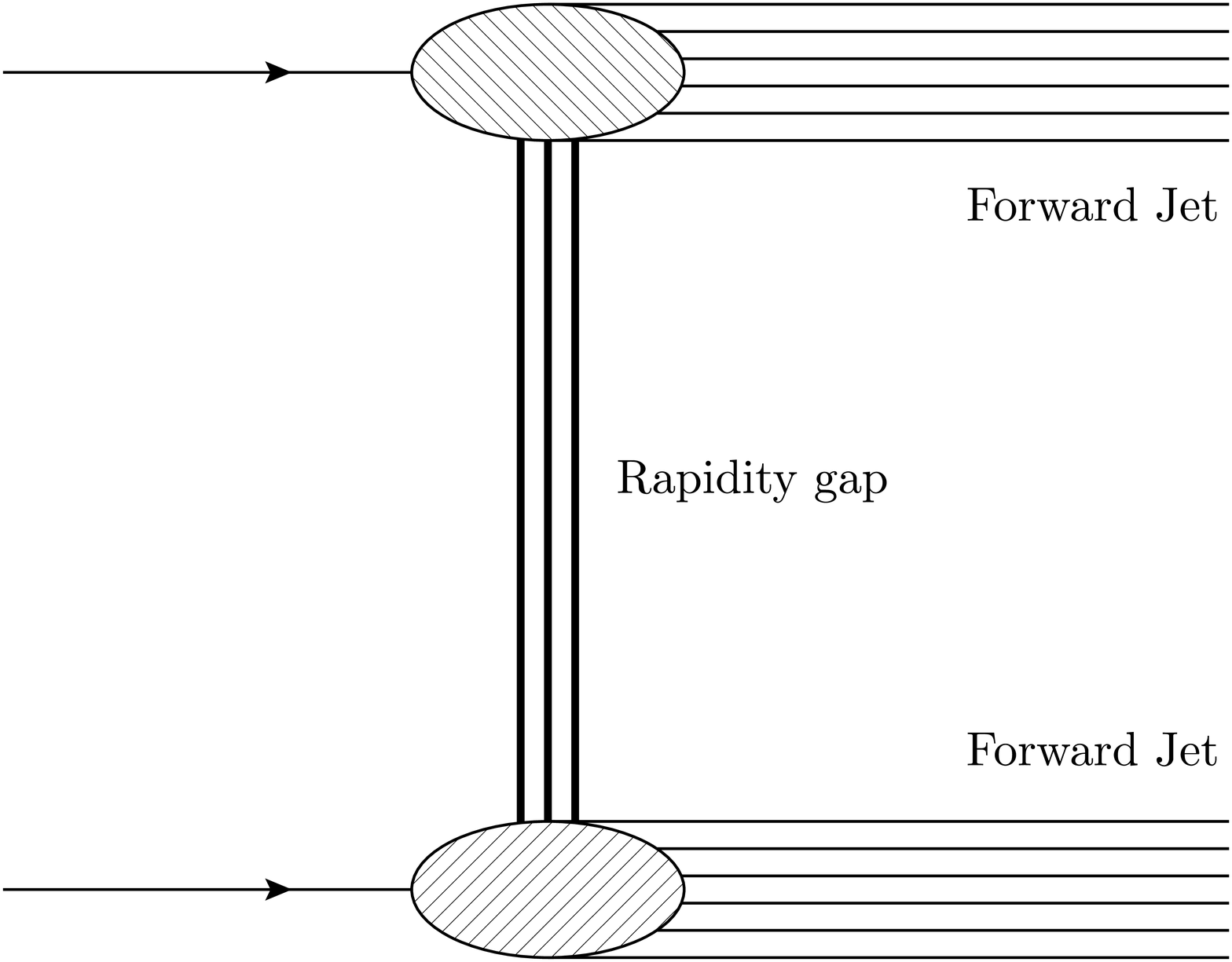} 
\end{minipage}
\end{centreFigure}
\begin{center}
a)\hspace*{2in} b)
\caption[]{Production processes for events with rapidity gaps.  a) shows the production by vector boson fusion of a Higgs boson and its decay to a pair of leptons.  b) shows the diffractive exchange of a pomeron.  In both cases there is no colour exchanged between the protons and a pair of forward jets are produced.}\label{fig:feynDiags}
\end{center}
\end{figure}

 Figure \ref{fig:feynDiags} shows the rapidity gap production in both the vector boson fusion process, which is not examined in this paper, and the diffractive production of hard jets, which is studied here.  The high cross section for hard diffraction allows gaps between jets to be studied with relatively  little data, and what is learnt can be applied to a later analysis of the vector boson fusion process.  However, the underlying event, that is multiple parton interactions between the same pair of protons, typically produces radiation throughout the whole detector, including in the gap region.    If a ``clean'' rapidity gap is required, only a small fraction (around 0.1,
depending on the model) of the original colour-singlet events (those with a very soft underlying event) can be retained.This could make it difficult or even impossible to distinguish colour connection di-jets (referred to as QCD di-jets from now on) from colour singlet di-jets.  Pile-up due to a high proton density in the beam bunch has a similar effect, bringing the efficiency for finding an empty gap almost to zero.  Clearly a better strategy for identifying gaps is needed.

The problem of separating the colour connection effect from the underlying event or pile up is one of separating features of differing physical size in the event.  As already stated, the underlying event fills the whole detector with radiation at all $\eta$.  Colour connection effects are smaller in size than the spread of radiation from the underlying event.  Colour connection effects are approximately the size of the jet-jet or jet-beamline interval.  Hadronisation and showering effects can be expected to be of a similar size to the colour connection effects.  The hard jets will be smaller still, with a radius of $R\simeq0.5$ and there may be jets originating from softer partons with $R$ as small as $0.1$.

\section{Monte Carlo Sample and Radiation Patterns}

Herwig \cite{Corcella:2002jc} was used to generate a sample of hadron-level QCD di-jet events (for the background) and diffractive colour singlet exchange events (for the signal).  The colour singlet exchange process in Herwig was adapted according to the prescription in \cite{Cox:1999dw} using code obtained from the authors.  In order to understand the effect of the underlying event we produced samples of QCD and colour singlet events both with and without the Jimmy \cite{Butterworth:1996zw} underlying event model turned on.  The analysis was run inside the Rivet \cite{Rivet} Monte Carlo analysis package, which provides an interface between the generator, the event record and the jet finding algorithm.

In order to apply an event selection we run the KT jet algorithm \cite{Cacciari:2005hq} using an R parameter of 0.7.  Events are chosen in which the two leading jets both have transverse energy, $E_{T}$, above 30~GeV, are separated by an $\eta$ interval of $\Delta\eta > 4 $ and are both within $\left|\eta\right|<5$.  No requirement is made on the absence or presence of radiation in the $\eta$ interval between the two leading jets.  We shall refer to the harder of the two jets as the hardest jet and the softer of the two jets as the softest jet.  To provide an input for the Fourier transform that is guaranteed to be safe against both soft and collinear divergences, we run the KT algorithm a second time with a much smaller R parameter of 0.1 and a minimum $E_{T}$ cut of 1~GeV.  Such small jets are also approximately the size of a calorimeter tower.

\begin{figure}
\begin{centreFigure}
\begin{minipage}[c]{0.45\linewidth}
\begin{overpic}[width=\linewidth]{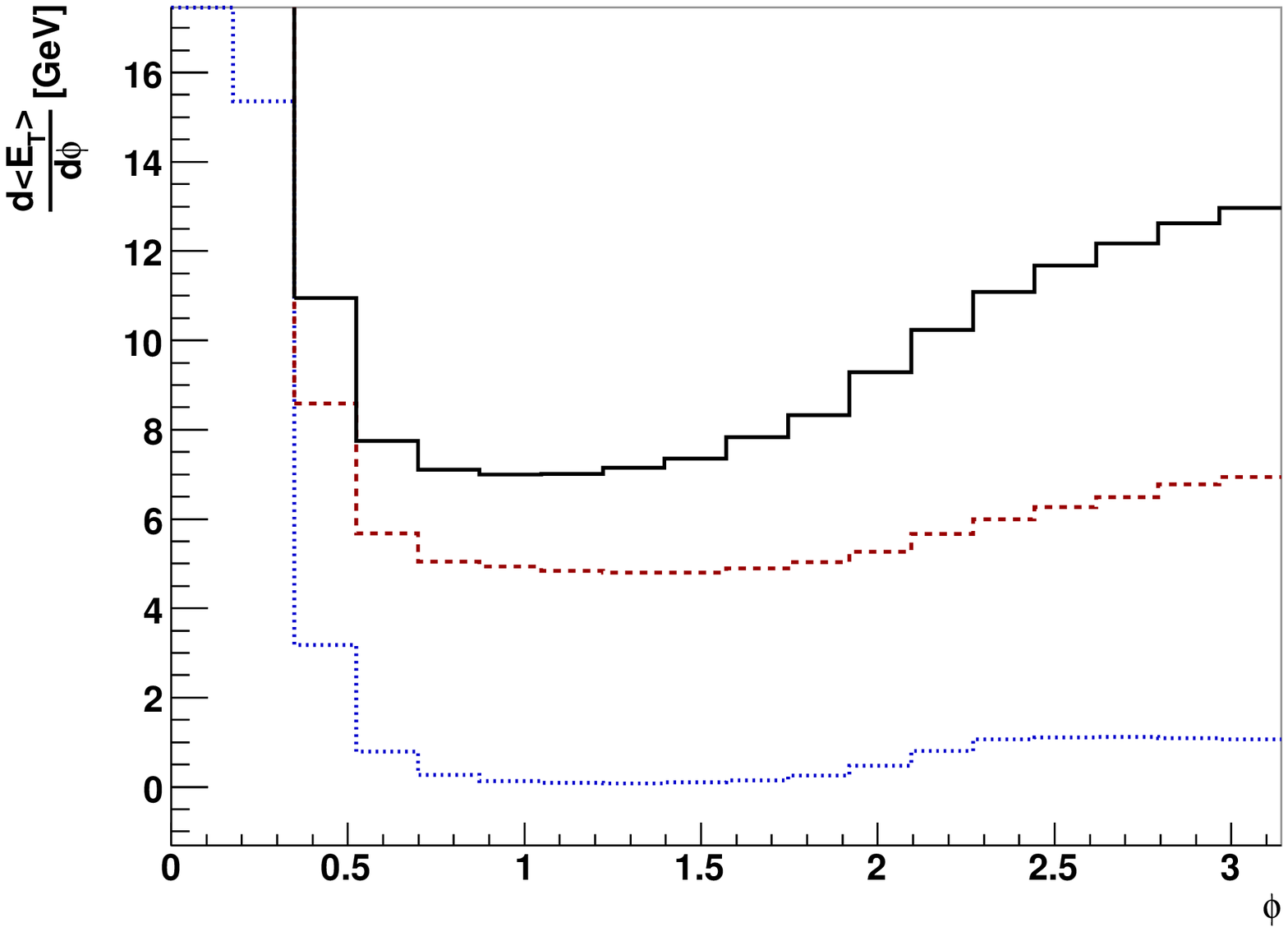}
\put(85,54){a)}
\end{overpic}
\end{minipage}
\begin{minipage}[c]{0.35\linewidth}
\begin{overpic}[width=\linewidth]{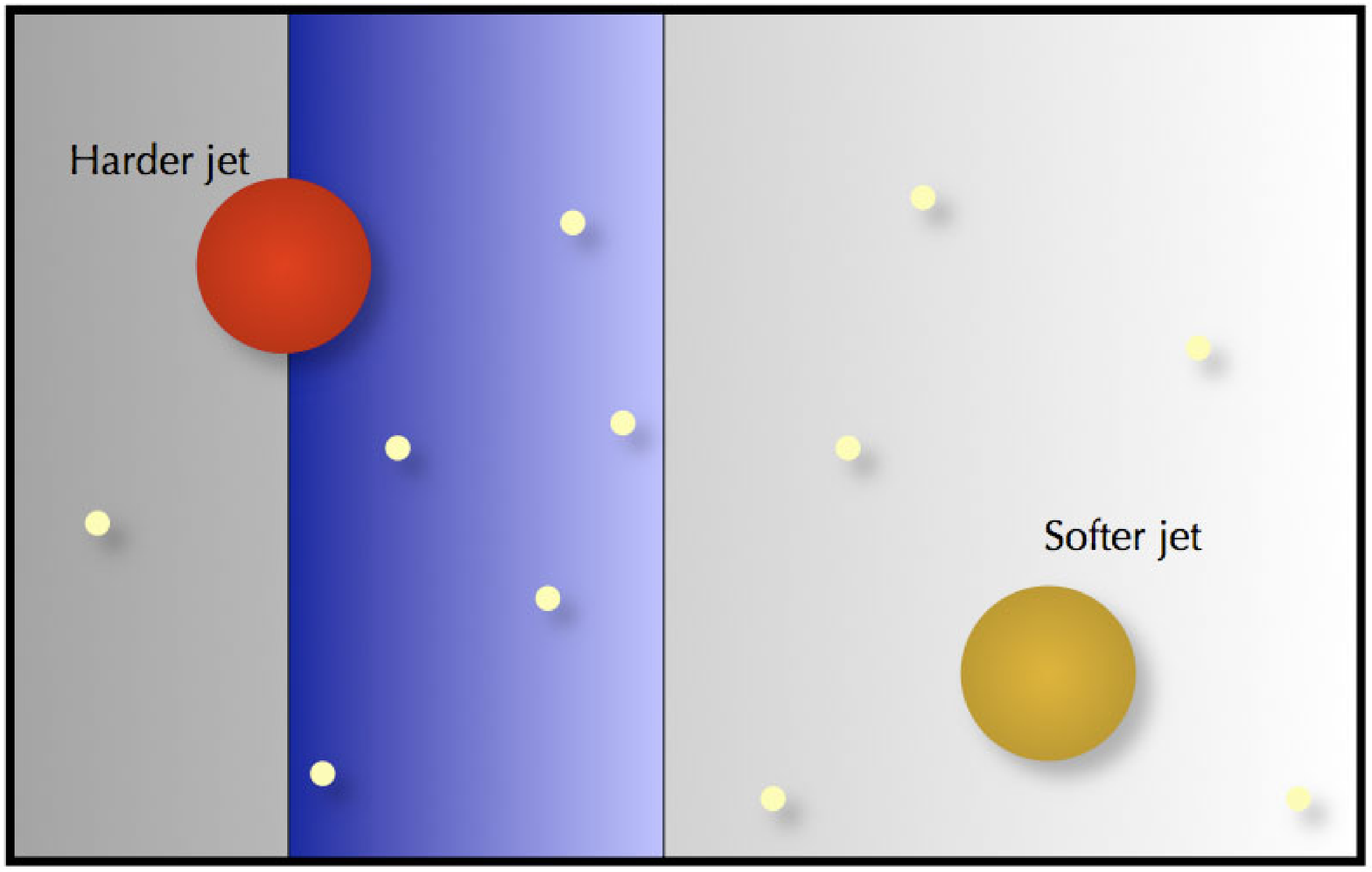}

\put(85, 60){b)}
\put(50, 5){$\eta$}
\put(99, 40){$\phi$}

\end{overpic}
\end{minipage}
\end{centreFigure}
\begin{center}
\caption[The $E_{T}$ flow in $\phi$]{The flow of $E_{T}$ Vs. $\phi$ in the half of the gap nearest the hardest jet.  a) shows that out-of-jet radiation from the softer jet causes a rise towards $\phi=\pi$.  The blue dotted line shows di-jet events produced from colour singlet exchange without underlying event, the red dashed line shows the effect of turning the underlying event on and the solid black line shows colour connected QCD di-jet events.  The y-axis has been truncated in order to better show the inter-jet radiation at $\phi>0.5$ rather than the hardest jet, which lies at $\phi=0$. b) shows the co-ordinate system and the region, shaded in blue, used for the $E_{T}$ flow in a).}\label{fig:ETFlow}

\end{center}

\end{figure}

The effect on the inter-jet radiation of turning first the underlying event and then the colour connection on can be seen in figure \ref{fig:ETFlow}, which shows the  $E_{T}$ flow against the distance $\phi$ from the hardest jet.  In figure \ref{fig:ETFlow}, and for this article \emph{only} in figure \ref{fig:ETFlow}, the $\eta$ interval between the two leading jets has been divided in two and we plot the $\phi$ distribution of radiation, weighted by $E_T$, in the region between the centre of the hardest jet and the middle of the gap. This is shown as the blue shaded region in figure \ref{fig:ETFlow}b) The hardest jet is at $\phi=0$, which leads to a large peak (off the scale in this plot to allow smaller features to be seen).  Away from the hardest jet there is a rise in $E_{T}$ towards $\phi=\pi$.  The rise is caused by showering and hadronisation from the softer jet on the opposite side of the $\eta$ interval.  This is despite the fact that the softer jet is itself not present in figure  \ref{fig:ETFlow}.  The three lines show the effect of colour singlet exchange without underlying event (blue dots), turning the underlying event on (red dashes) and a colour connection between the two leading jets (solid black).  Underlying event adds radiation uniformly in $\phi$ and colour connection effects increase the rise of $E_{T}$ with $\phi$ due to an enhancement of showering and hadronisation.  The aim of the Fourier analysis will be to identify features such as these without looking only at a localised region of a detector.

\section{Fourier decomposition}

Unlike the case of the CMB, hadron collision detectors do not have complete $4\pi$ coverage of events; they are typically bounded and periodic in the $\phi$ direction but not in the $\eta$ direction.  Further, the cylindrical co-ordinate system together with the additive nature of rapidity under longitudinal boosts is more suited to cylindrical rather than spherical harmonics.   The simplest Fourier transform that can be made, and therefore the one that we investigate first, is the discrete one dimensional Fourier transform of the $E_{T}$ distribution in $\phi$. 

The $\phi$ co-ordinate is defined such that the centroid of the hardest jet lies at $\phi=0$.  The positive $\phi$ direction is defined such that the softer jet lies between $\phi=0$ and $\phi=\pi$.  The co-ordinate system is divided into a grid of $N=32$ bins in $\phi$, each bin covering the region $-5<\eta<5$.   Note that the hardest two  jets are used only to set the position of $\phi=0$ and the direction of increasing $\phi$.   The $E_{T}$ sum, $E_{T}\left(\phi_{l}\right)$, is calculated in each bin from the KT 0.1 clusters.   All activity above the 1~GeV threshold for the KT 0.1 jets is included in the summation.  The centre of the $l=0$ bin is aligned with $\phi=0$, the centre of the hardest jet.  Notice that the input to the Fourier decomposition is 32 real $E_{T}\left(\phi_{n}\right)$ and the output is 32 complex coefficients.  Performing the Fourier transform therefore seemingly 
doubles the number of degrees of freedom.  However, there is a symmetry between the $n^{th}$ and $\left( N-n\right)^{th}$ coefficient such that

\begin{eqnarray}
C_{n} & = & \frac{1}{\sqrt{N}}\sum_{l=0}^{N}E_{T}\left(\phi_{l}\right) e^{i n\phi_{l}} \nonumber\\
C_{N-n} & = &\frac{1}{\sqrt{N}}\sum_{l=0}^{N}E_{T}\left(\phi_{l}\right)  e^{-i n \phi_{l}} e^{i N \frac{2\pi l}{N}}\nonumber\\
C_{N-n} & = & C_{n}^{\star}
\end{eqnarray}
so there are only 16 independent complex coefficients in the output, matching the amount of information in the input distribution.

\begin{center}
\begin{figure}
\begin{center}
\includegraphics[width=0.85\columnwidth, clip=true]{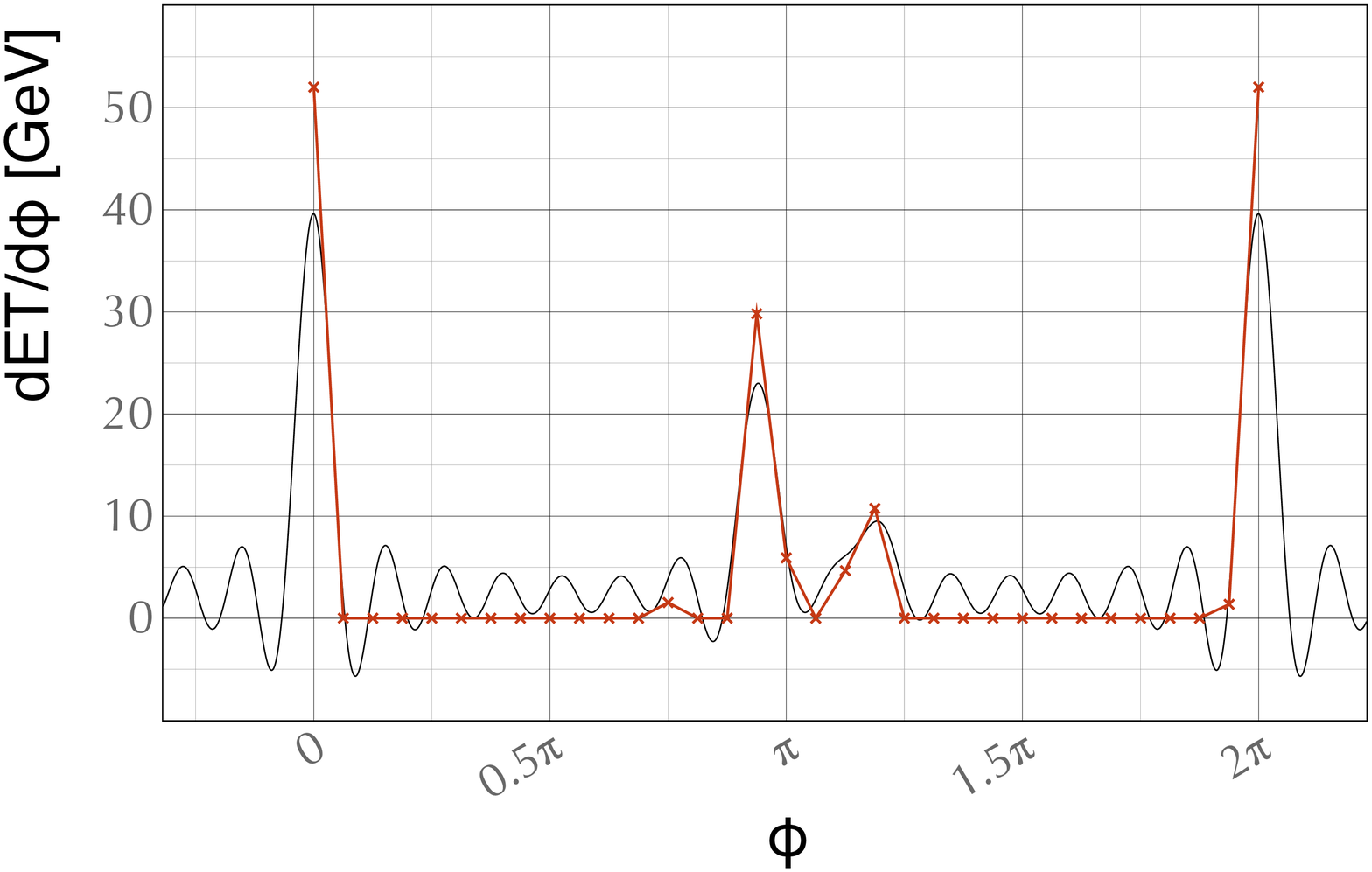}\raisebox{0.3\columnwidth}{a)}
\includegraphics[width=0.85\columnwidth, clip=true]{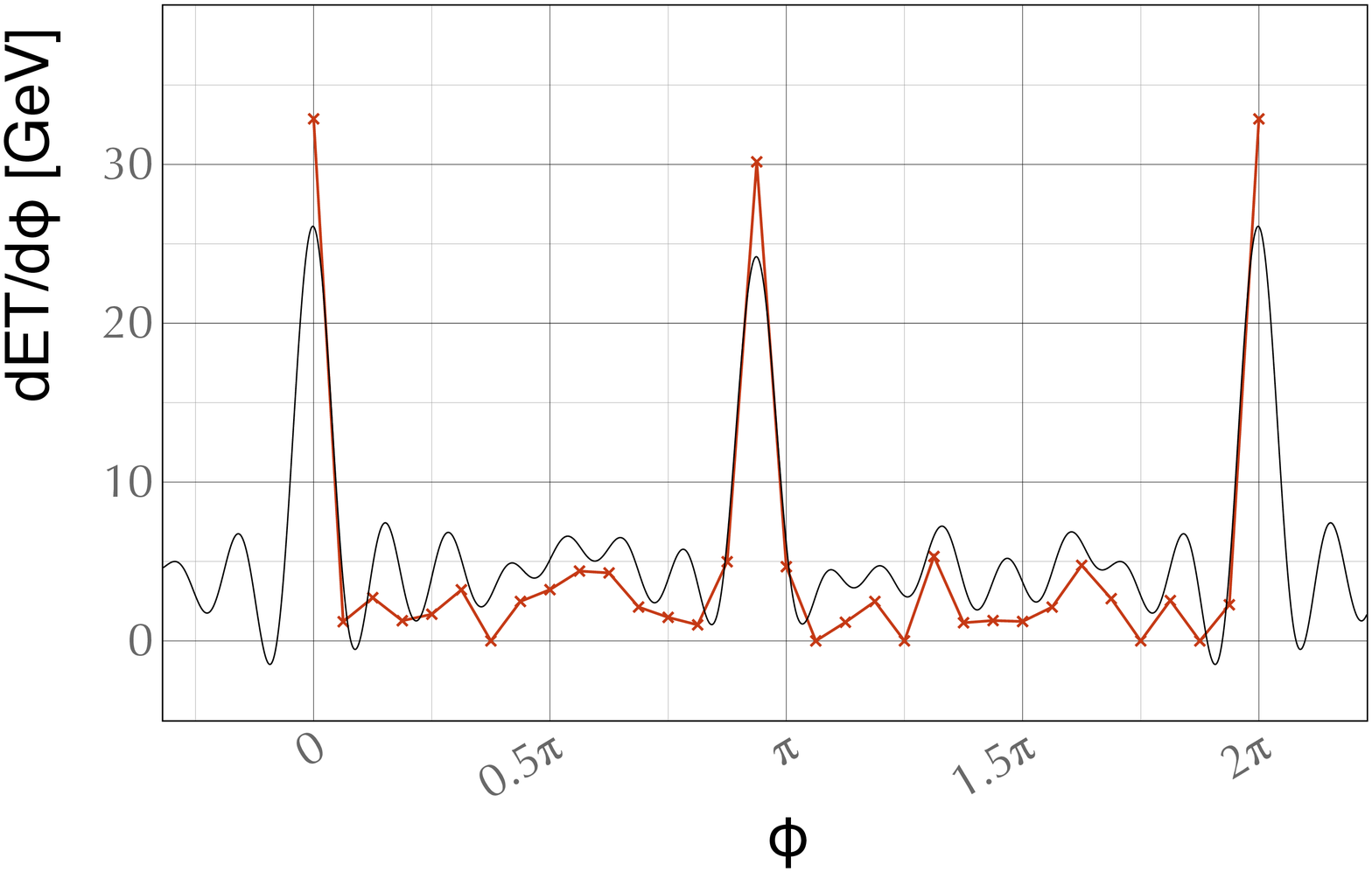}\raisebox{0.3\columnwidth}{b)}
\end{center}
\caption[Inverse Fourier transform overlaid on input data]{Inverse Fourier transform of the $N=32$ coefficients (black curve) compared to the input data (red line), both for the same single event. a) (top) shows a colour singlet exchange event without underlying event and contains almost no activity away from the leading jets.  b) (bottom) is a QCD di-jet event with underlying event and shows activity away from the leading jets.  In both cases the small features of radiation between the jets are present in the Inverse function.  Note that the Black curve has been scaled down by a quarter in order to overlay on the input $E_{T}$.}\label{fig:reverse}

\end{figure}
\end{center}

We use the GNU Scientific Library (GSL) \cite{GSL} to provide fast one dimensional Fourier transform routines.  Note that these routines perform best if the number of input data bins can be written as a power of $2$, $N = 2^{\hat{N}}$,  with $\hat{N}$ an integer.  In order to verify that the transformation has been performed correctly we take the output coefficients for a single event, reverse the transformation and overlay the resulting function on the input set of $E_{T}$ bins.  The result in figure \ref{fig:reverse} shows that the curve produced by  the Fourier coefficients matches the input distribution.  The match is good for a colour singlet exchange event that contains very little activity other than the di-jets and a QCD di-jet event that includes activity outside of the leading jets.

\section{Fourier Analysis Distributions}

\begin{figure}
\begin{centreFigure}
\begin{minipage}[b]{0.38\columnwidth}
\begin{overpic}[width=\linewidth]{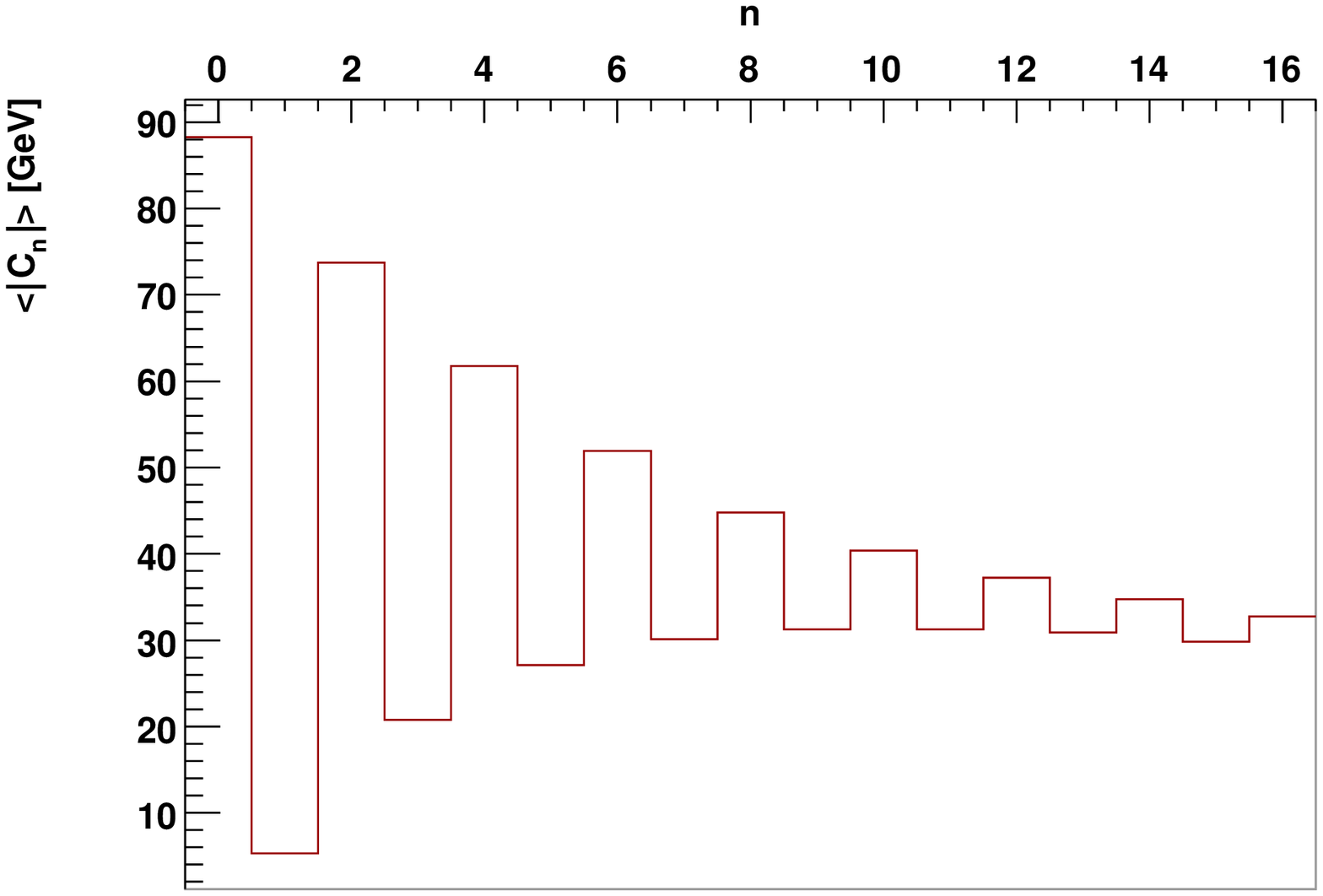}
\put(80,50){a)}
\end{overpic}

\end{minipage}
\begin{minipage}[b]{0.38\linewidth}
\begin{overpic}[width=\linewidth]{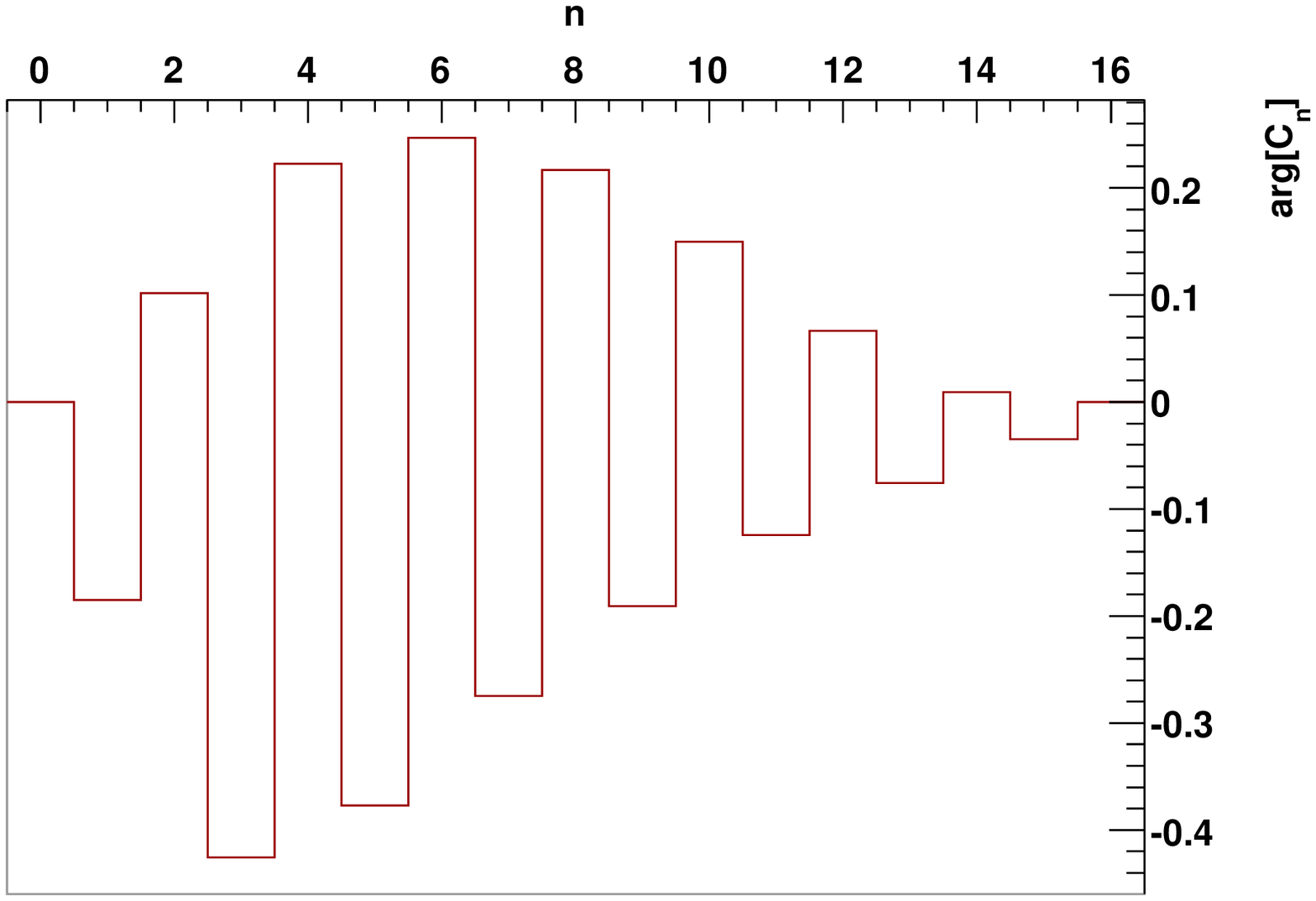}
\put(70,20){b)}
\end{overpic}

\end{minipage}\\
\begin{minipage}[b]{0.38\linewidth}
\begin{overpic}[width=\linewidth]{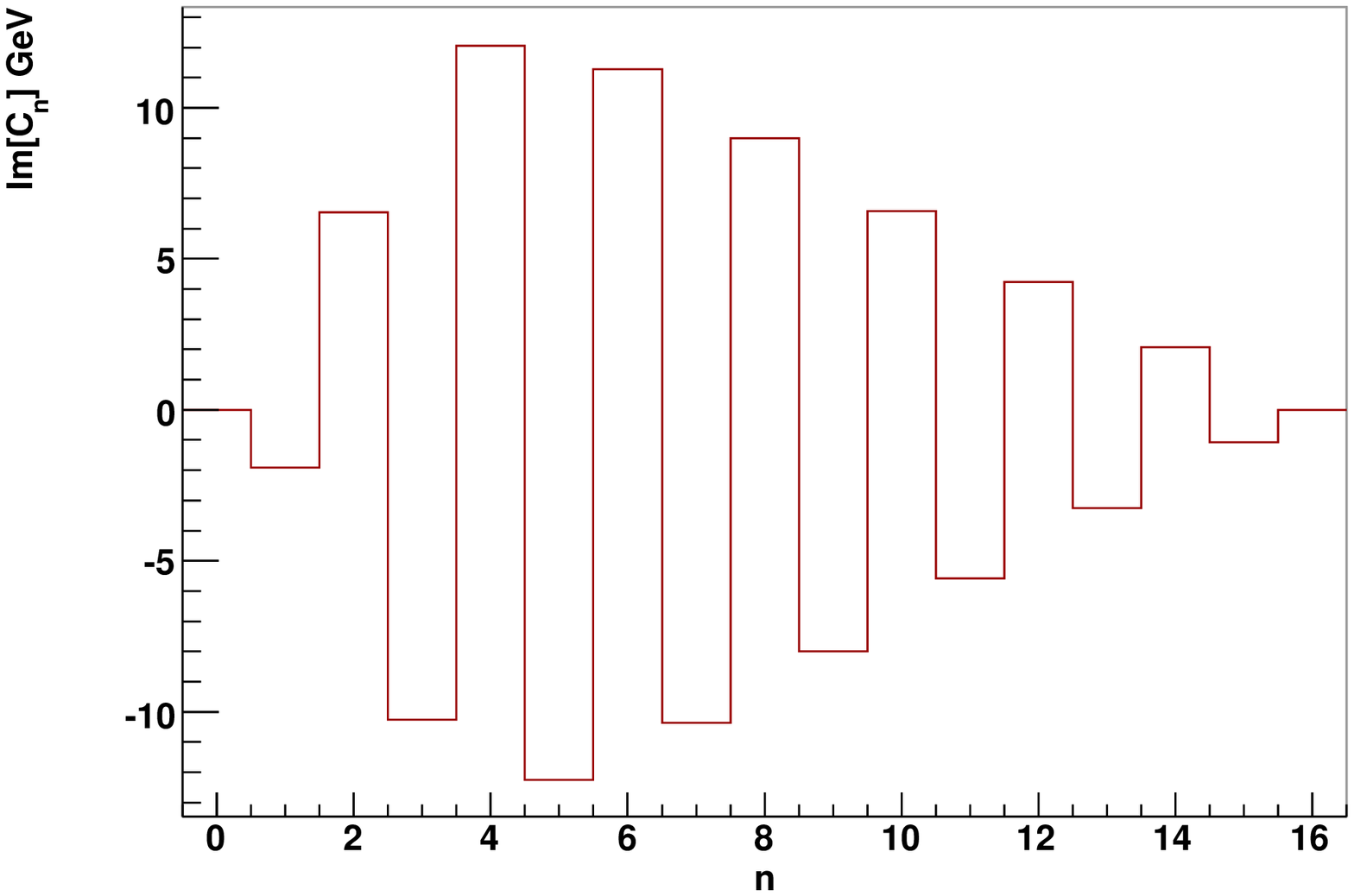}
\put(80,50){d)}
\end{overpic}
\end{minipage}
\begin{minipage}[b]{0.38\linewidth}
\begin{overpic}[width=\linewidth]{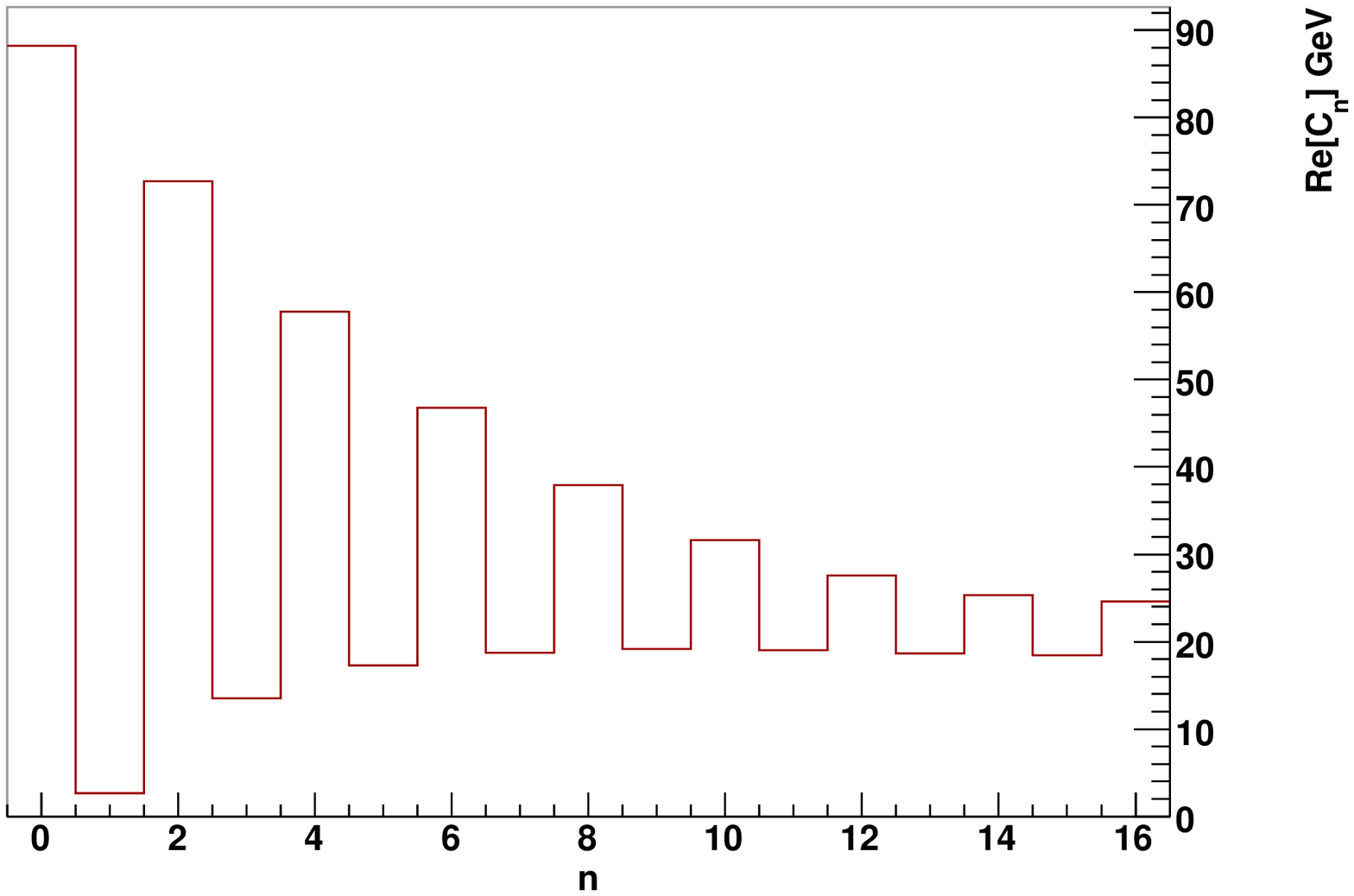}
\put(70,50){c)}

\end{overpic}
\end{minipage}
\end{centreFigure}
\begin{center}
\caption[The mean magnitude, phase, real part and imaginary part of the first sixteen coefficients.]{Clockwise from top left: the mean a) magnitude, b) phase, c) real part and d) imaginary part of the first sixteen coefficients for colour singlet exchange events without underlying event.}\label{fig:CSEMag}
\end{center}
\end{figure}

\begin{figure}
\begin{centreFigure}

\includegraphics[width=0.5\paperwidth]{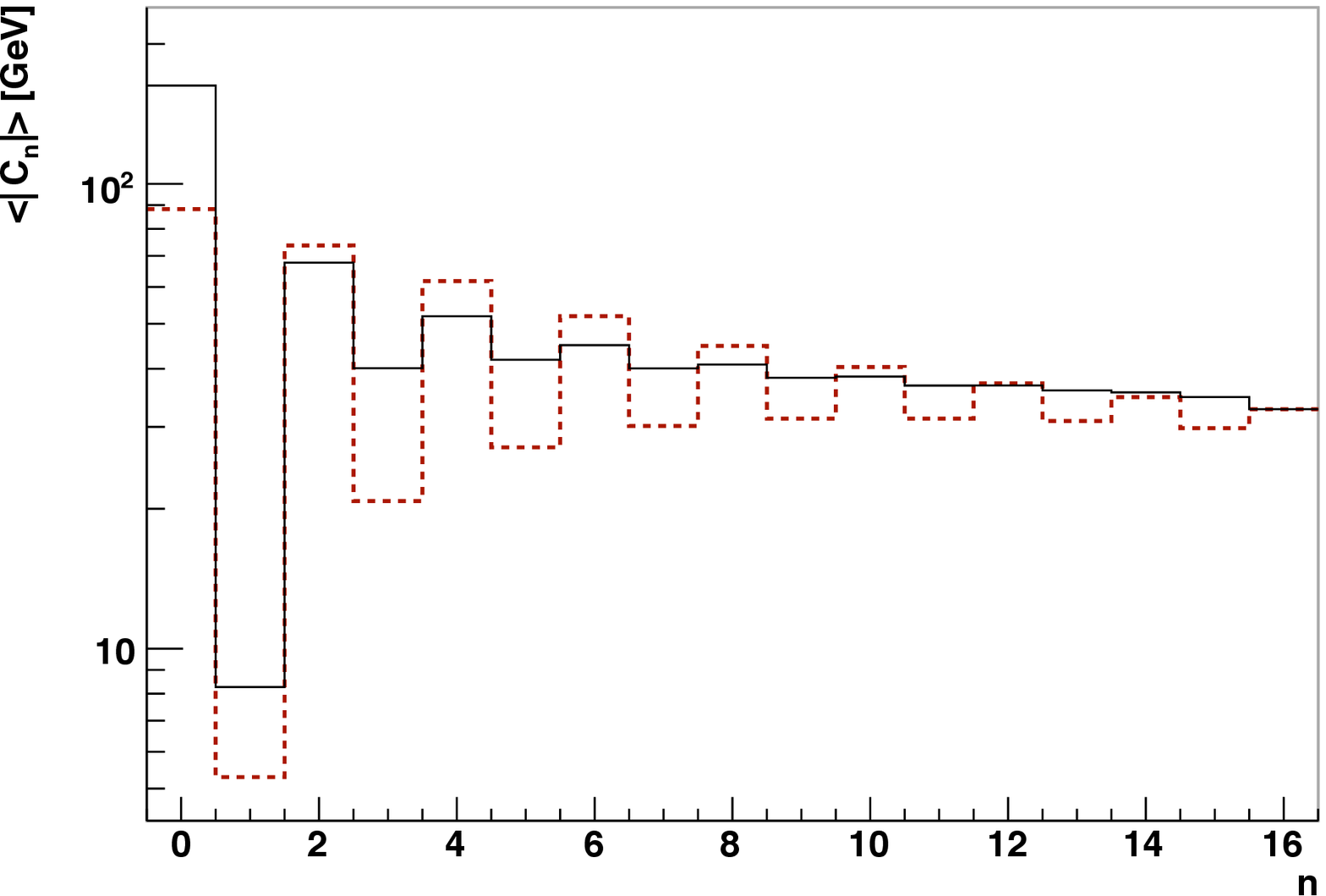}
\end{centreFigure}
\begin{center}
\caption[The average magnitude of the Fourier coefficients for QCD di-jets with underlying event]{The average magnitude of the first 16 Fourier coefficients for QCD di-jets with underlying event (black solid line).  Compared to the colour singlet jets without underlying event (red dashed line) the events with colour flow show less suppression of the small $n$ odd coefficients.}\label{fig:QCDMag}
\end{center}

\end{figure}

We apply the Fourier transform to a sample of 15 million colour singlet events without underlying event, which is the most purely di-jet-like  sample considered. The average magnitude, real part, imaginary part and phase of the first sixteen coefficients, $C_{n}$, are shown in figure \ref{fig:CSEMag}.  Both the magnitude and real part show a suppression of the odd coefficients, especially for the lower frequencies.  The $n^{th}$ coefficient corresponds to features in the event of size $R\simeq\pi/n$.  Since the phase of each event is chosen so that the hardest jet appears at $\phi=0$, the odd coefficients \emph{always} produce a trough at $\phi=\pi$, which is the approximate location of the softer jet.  Odd coefficients therefore represent features that deviate from the behaviour of back-to-back di-jets.  The small $n$ odd coefficients therefore correspond to non-di-jet like features that are large; in other words radiation between the jets.  Since this colour singlet sample contains very little radiation between jets, the odd coefficients are suppressed. This is confirmed in figure \ref{fig:QCDMag}, in which the average magnitudes of the coefficients for five million colour connected QCD di-jets with underlying event show a smaller suppression of the odd coefficients, indicating that these events are less di-jet like, as expected.

\begin{figure}
\begin{centreFigure}

\begin{overpic}[width=\linewidth]{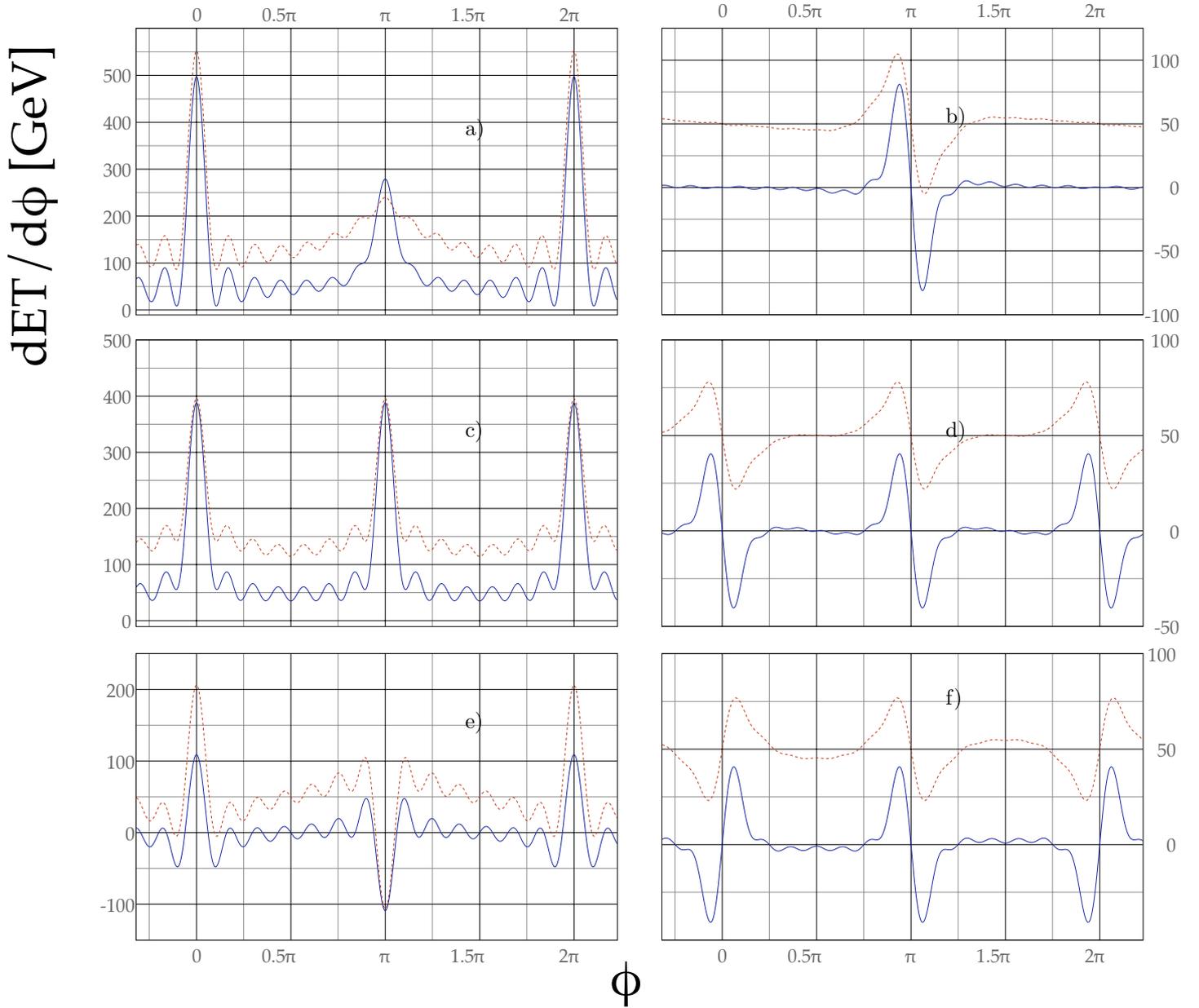}

\put(39, 75){a)}
\put(80, 76){b)}
\put(39, 49){c)}
\put(80, 49){d)}
\put(39, 24){e)}
\put(80, 26){f)}

\end{overpic}

\end{centreFigure}
\begin{center}
\caption[]{The reverse Fourier transform using the average coefficients over the colour singlet sample without underlying event (blue solid curve) and the QCD di-jets with underlying event (red dashed curve).  a) uses only the real part of the coefficients, b) only the imaginary part.  c) uses only the real even coefficients, d) uses only the imaginary even coefficients.  e) uses the real odd coefficients and f) uses the imaginary odd coefficients.  For all but a) and c) the red dashed curve has been raised by 50~GeV in order to separate it from the blue solid curve.}\label{fig:EvenOdd}
\end{center}
\end{figure}

The meaning of the separate real and imaginary or even and odd coefficients can best be interpreted using figure \ref{fig:EvenOdd}, in which the inverse Fourier transform is performed on the average coefficient over the event samples for both the colour singlet (no underlying event) and QCD (with underlying event) di-jets.  The reverse transformation is performed separately for the real, imaginary, even and odd coefficients.  For these di-jet events the real even part (c) is largely responsible for recreating the hardest jet.  Using only the real even coefficients one obtains exact di-jets.  The real odd part (e) describes the difference in $E_{T}$ between the two leading jets.  Note the trough either side of $\phi=0$, which serves to narrow the hardest jet.  The same feature is inverted at $\phi=\pi$ and broadens the softer jet.   The imaginary part (b) contains no information about the harder jet, but is responsible for the shift of the softer jet away from $\phi=\pi$ and an additional broadening. 

\begin{figure}
\begin{centreFigure}
\begin{minipage}[b]{0.4\columnwidth}
\begin{overpic}[width=\linewidth]{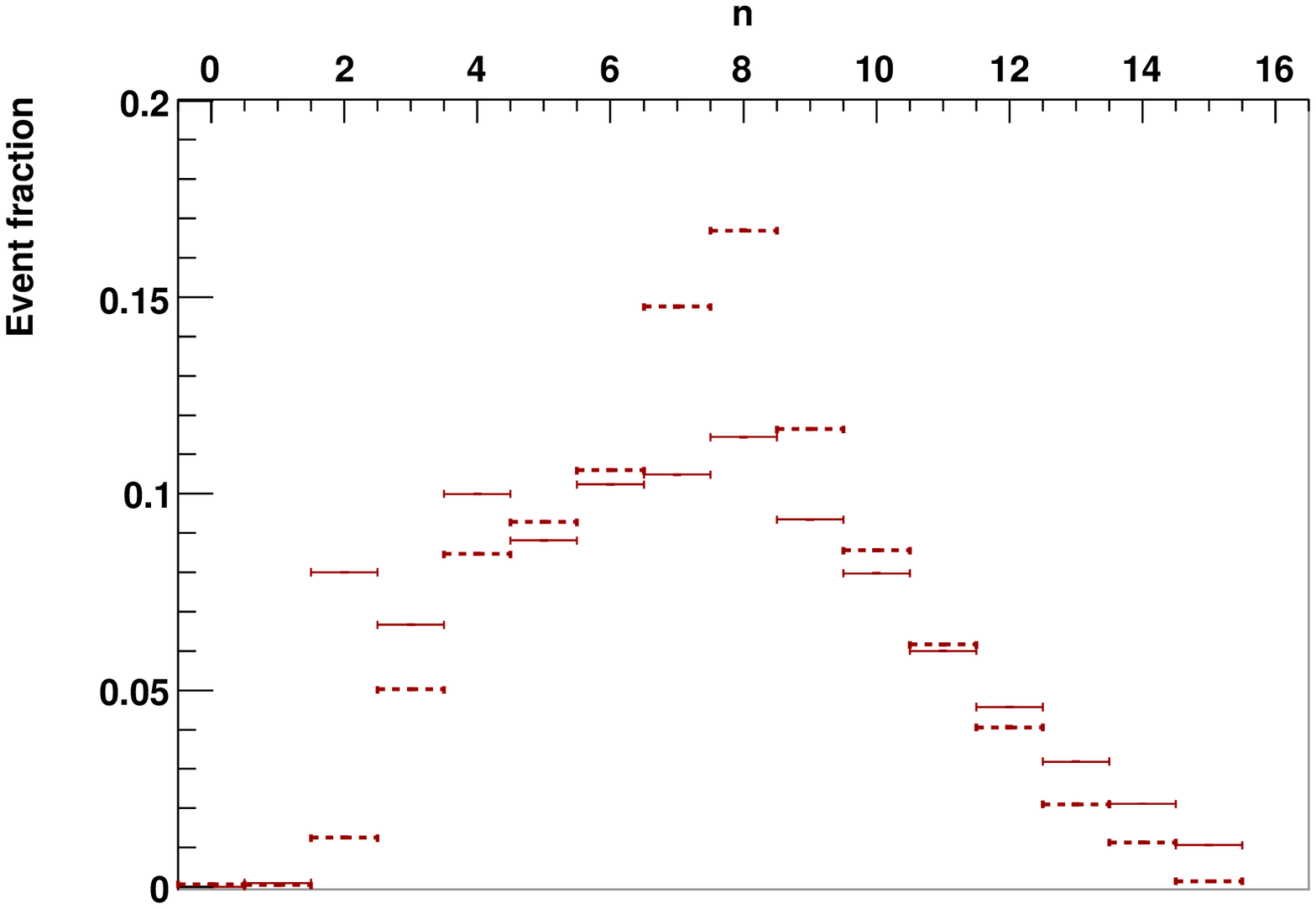}
\put(80,50){a)}
\end{overpic}
\end{minipage}
\begin{minipage}[b]{0.4\columnwidth}

\begin{overpic}[width=\linewidth]{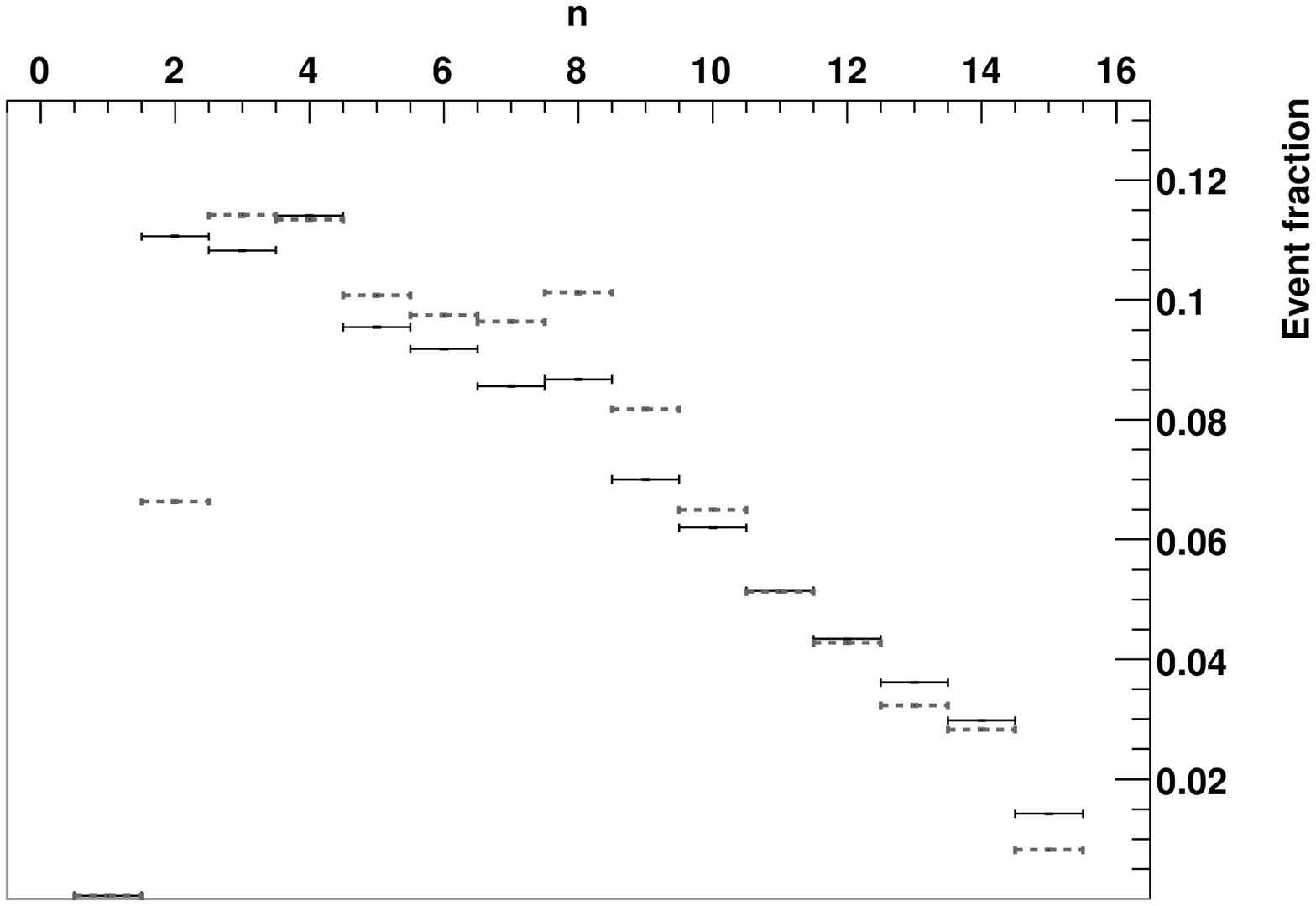}
\put(70,50){b)}
\end{overpic}

\end{minipage}
\begin{minipage}[b]{0.4\columnwidth}

\begin{overpic}[width=\linewidth]{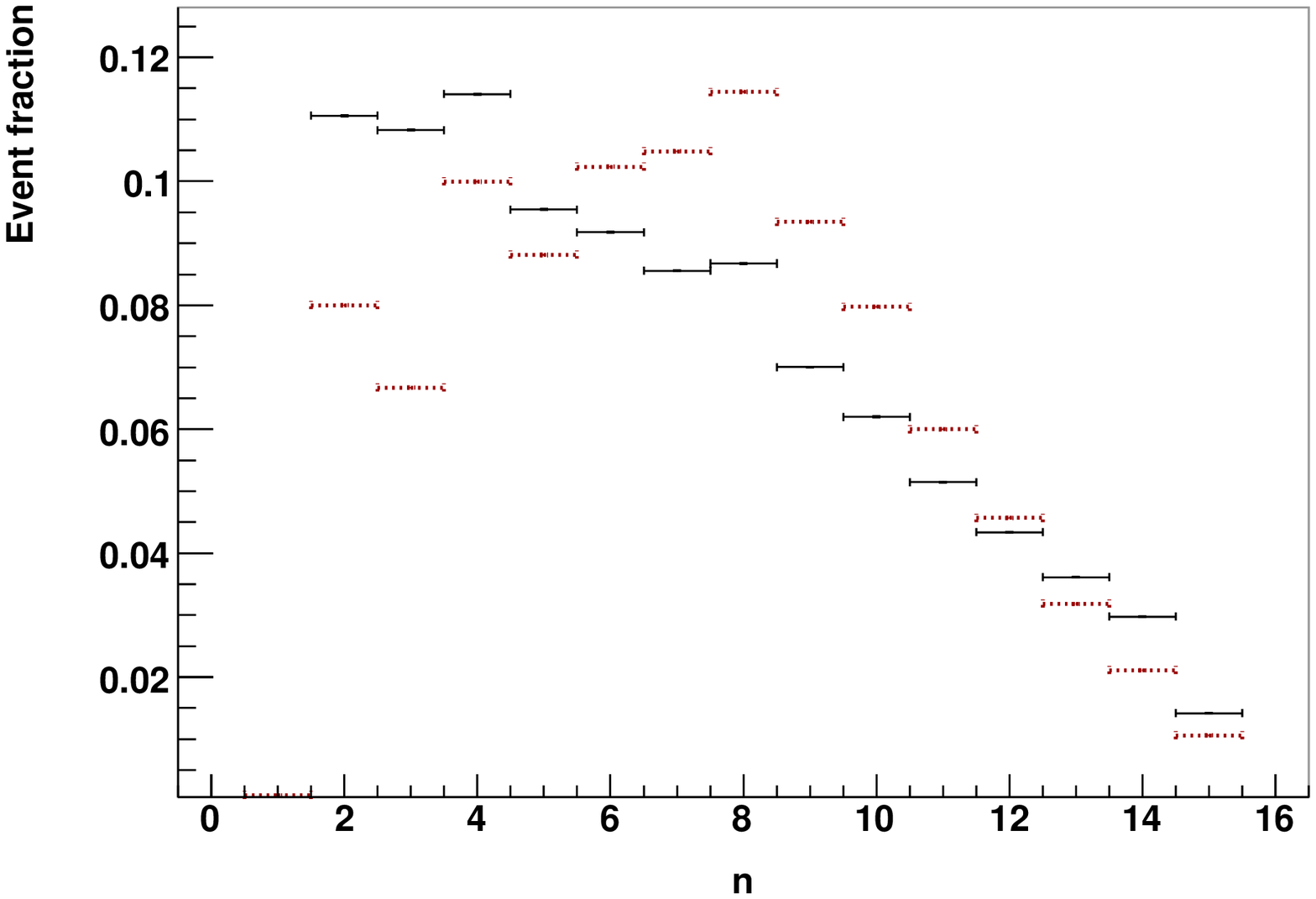}
\put(80,50){c)}
\end{overpic}

\end{minipage}
\hspace*{0.02\columnwidth}\begin{minipage}[b]{0.32\columnwidth}
\begin{center}
\caption[The coefficient with the largest imaginary part in each event.]{The coefficient with the largest imaginary part in each event.  a) shows colour singlet events with (solid) and without (dashed) underlying event.  Top right shows QCD di-jets with (solid) and without (dashed) underlying event and bottom left compares QCD di jets (solid-black) with those from colour singlet exchange (red-dotted) - both with underlying event.}\label{fig:largestIm}
\end{center}
\end{minipage}\hspace*{0.06\columnwidth}
\end{centreFigure}

\end{figure}

The imaginary part of the coefficients in figure \ref{fig:CSEMag} shows a peak around $n=4-6$, which corresponds to features of size $R\simeq0.6$, the typical size of the hadronic jets.  The choice of phase means that the imaginary components correspond to radiation away from the centroid of the hardest jet.  Such radiation could either be the softer jet or radiation between the jets.  Figure \ref{fig:largestIm} shows the distribution of the largest imaginary coefficient in each event, comparing colour singlet events with and without underlying event and QCD di-jets.  The colour singlet sample shows a peak in the largest imaginary coefficient around $n=8$ (again, roughly the size of a jet), which turns into a shoulder once the underlying event is present.  The underlying event affects the small $n$ coefficients more than the medium to large $n$.  The QCD di-jets show a peak around $n=3$ because the larger scale radiation between jets is more significant in that sample.  There  is a bump in the distribution around $n=8$, corresponding to the jet-like features that are also present (and more dominant) in the colour singlet sample.  Turning on the underlying event favours lower $n$ coefficients and flattens the bump around $n=8$.  Overall, comparing QCD with colour singlet di-jets shows that, even in the presence of underlying event, the colour singlet sample has a broad peak around $n=8$, indicating that those events are more back-to-back di-jet like.  On the other hand, the QCD sample is dominated more by the lower coefficients that arise from the inter-jet radiation caused by  colour connection effects, hadronisation and showering.  Although the bump is less clear, it is nevertheless visible even in the presence of underlying event, showing the potential of a Fourier analysis.

Note the relative depletion of the $3^{rd}$ and (to a lesser extent) $5^{th}$ coefficients in the colour singlet sample with underlying event.  The $n=3$ and $n=5$ coefficients are expected to correspond to radiation between the jets and are at large enough $n$ that the underlying event is not completely dominant.  Thus the $n=3, 5$ coefficients may be sensitive to differences in colour connection effects, showering and hadronisation.

\begin{figure}
\begin{centreFigure}
\begin{minipage}[b]{0.4\columnwidth}

\begin{overpic}[width=\linewidth]{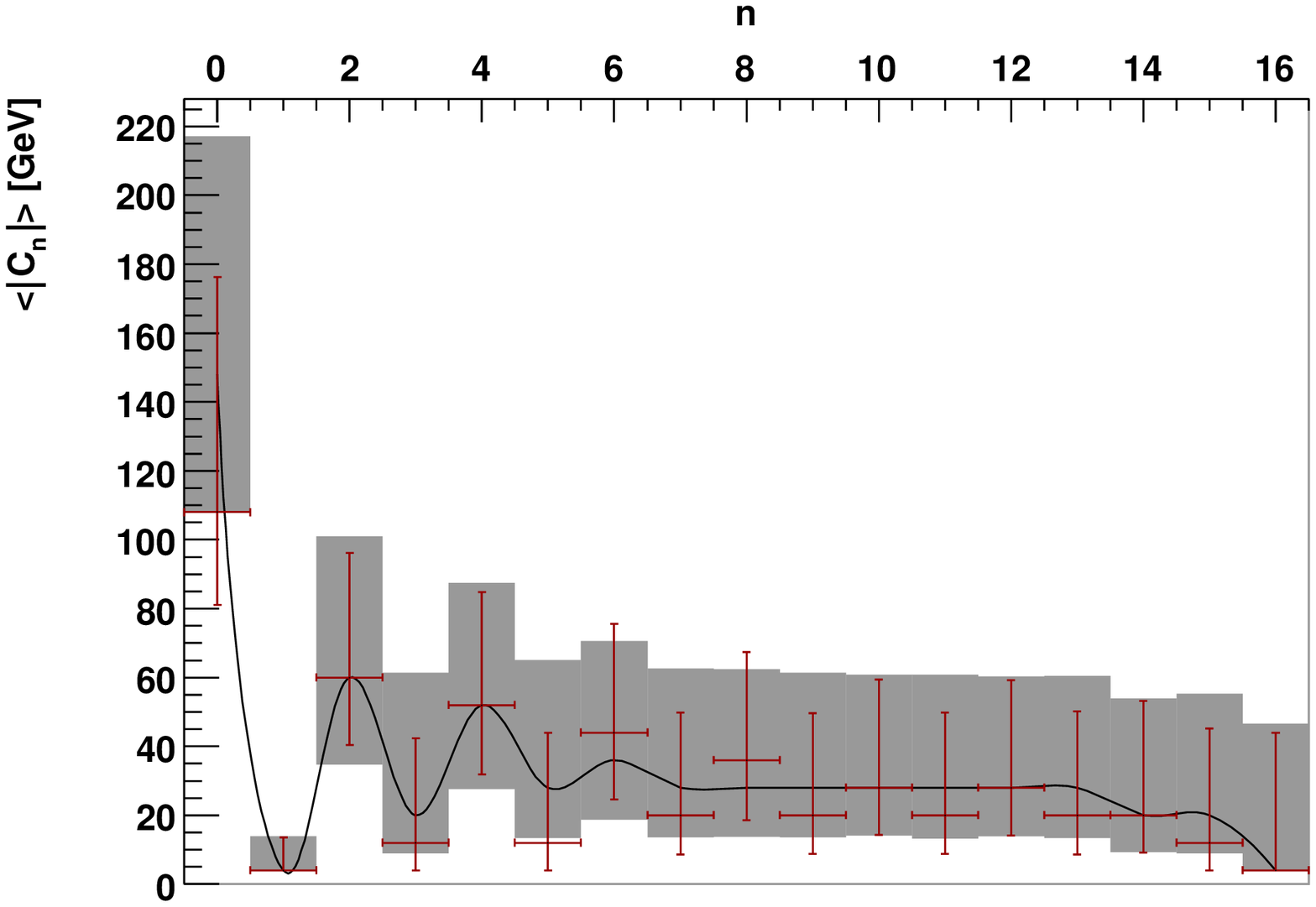}

\end{overpic}
\end{minipage}
\begin{minipage}[b]{0.4\columnwidth}
\begin{overpic}[width=\linewidth]{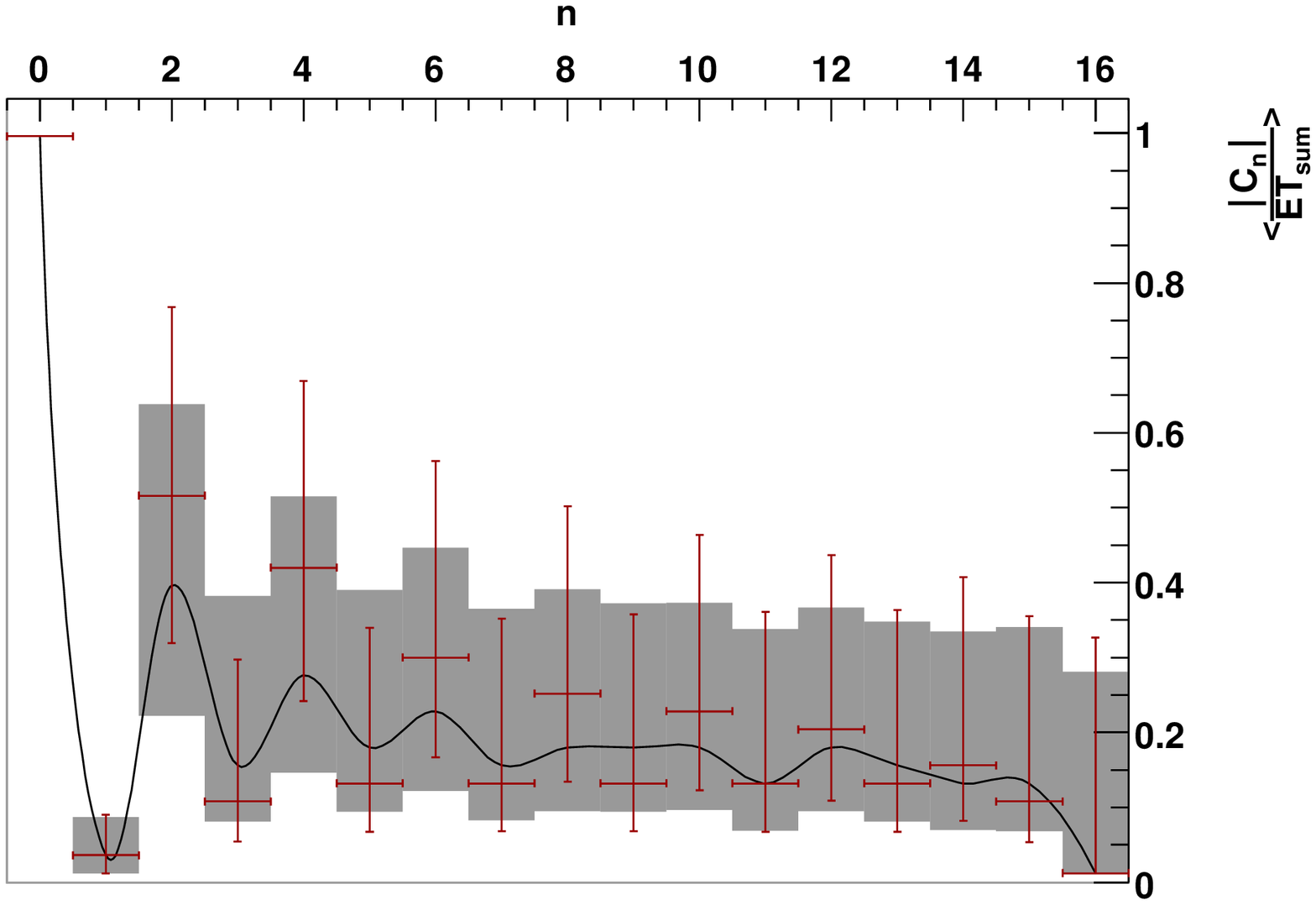}

\end{overpic}
\end{minipage}

\begin{minipage}[b]{0.4\columnwidth}
\begin{overpic}[width=\linewidth]{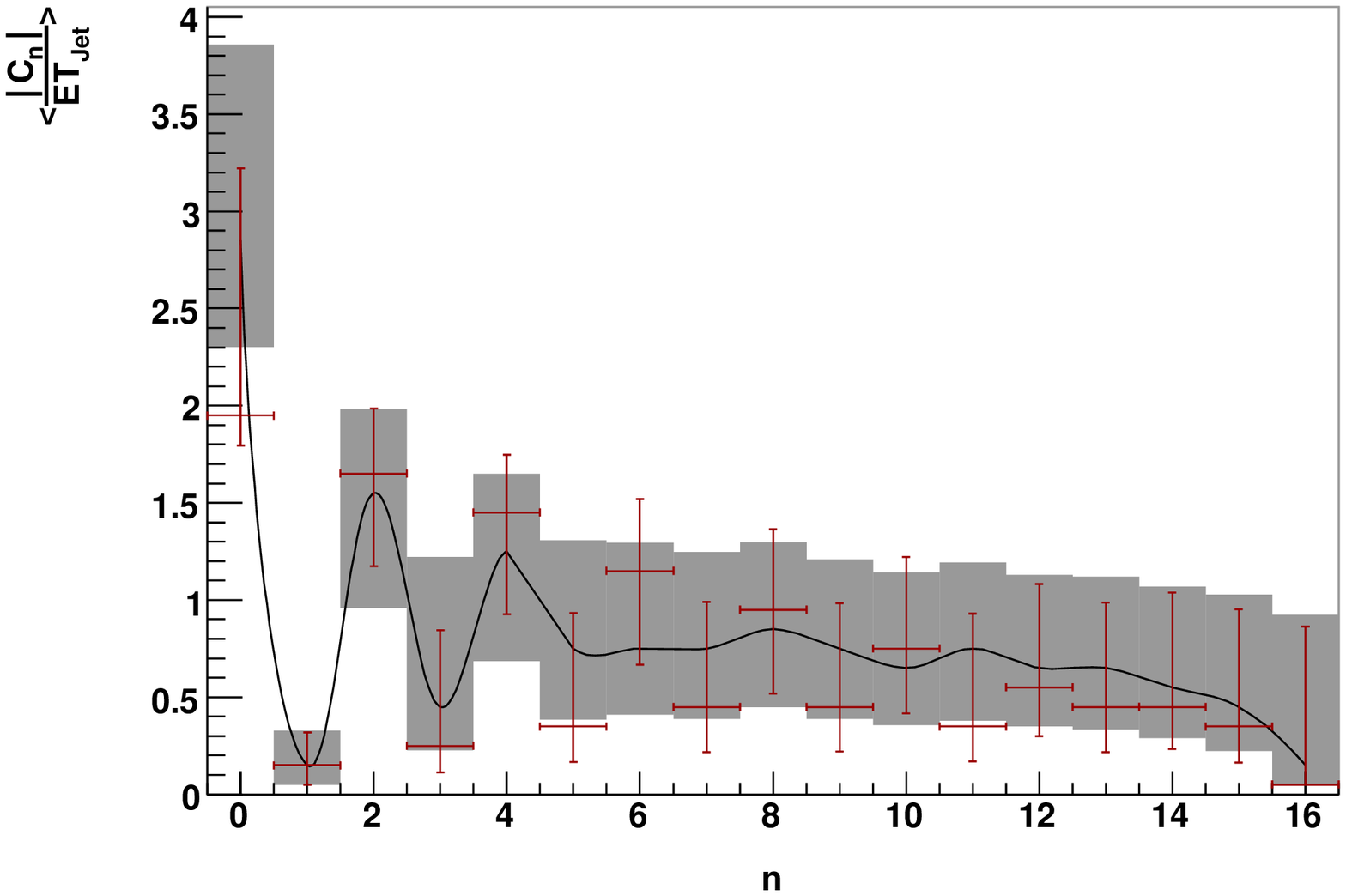}

\end{overpic}
\end{minipage}
\begin{minipage}[b]{0.4\columnwidth}
\caption[]{The spread of the coefficients of the Fourier transform.  The black curve shows the modal value for the coefficients from the QCD sample, the grey shaded region showing the spread about the mode.  The red crosses show the colour singlet exchange. a) Shows the spread un-normalised, b) shows the spread when the coefficients for each event have been normalised by the $E_{T}$ sum in the event and c) shows the effect of normalising each event by the $E_{T}$ of the leading jet.}\label{fig:spread}
\end{minipage}

\end{centreFigure}

\begin{center}

\end{center}

\end{figure}

In figure \ref{fig:spread} we again show the distribution of the magnitudes of each of the first sixteen coefficients, comparing QCD with colour singlet di-jets, both with underlying event.  The centre of the red cross or the black curve shows the modal value for each coefficient, with the band above and below showing the RMS spread on either side of this value.  The spread of each of the coefficients is due to both variations in the event shape and because there is a spread of leading jet $E_{T}$ values.  By dividing the $E_{T}$ sums in the grid that is input to the Fourier transform by either the total $E_{T}$ in the event or the $E_{T}$ of the leading jet one can attempt to narrow the spread of certain of the coefficients. Figure \ref{fig:spread}a) shows the coefficients without any additional normalisation, \ref{fig:spread}b) shows the same coefficients normalised by the total $E_{T}$ sum in the grid and \ref{fig:spread}c) shows the effect of normalising by the $E_{T}$ of the hardest jet.  Normalising by the total $E_{T}$ sum narrows the spread of the $n=0$ coefficient (in fact the $n=0$ coefficient is itself the $E_{T}$ sum), supporting the claim that the low coefficients are sensitive to the underlying event, which is present at all $\phi$.  The spread of the higher coefficients, however, appears to in fact be worsened by normalising to the total $E_{T}$.  On the other hand, normalising by the $E_{T}$ of the hardest jet does not change the spread of the coefficients very much, but it does enhance the separation between the alternating even and odd coefficients.  Compare, for example, the overlap between the $4^{th}$ and $5^{th}$ coefficients for colour singlet exchange in \ref{fig:spread}a) and \ref{fig:spread}c).   

\section{Conclusion}

Fourier analysis is a relatively unexplored tool for high energy physics that, based upon this first simple application, appears to be quite interesting.  We believe that a very promising area for its application is the study of the different scales present in fully-hadronic final states in a hadron collider.  In the best-case scenario a Fourier decomposition should be able to separate the different effects of hard scattering, showering, hadronisation and the underlying event, each one of them having a distinctive $\eta-\phi$ scale. 

In this article, we show the application of this idea, still in the one-dimensional form, to the analysis of di-jet events with large rapidity separation between the main jets. In particular we have shown the effect of the underlying event and specific features of colour singlet and colour octet exchange processes. Features that are otherwise difficult to extract in a traditional analysis emerge in a quite clear way.

We aim to extend the Fourier transform to a full two dimensional decomposition, and we hope the additional information thus preserved will render the analysis more powerful.  One key aspect of this method is that each event is aligned so that the hardest jet is at $\phi=0$, which  gives a common phase to each event.  Whilst this is simple to do for the one dimensional transformation in $\phi$, doing the same in the $\eta$ direction will restrict the $\eta$ position of the jets within the detector.

\section*{Acknowledgments}

We would like to thank Norbert Wermes for his useful input to this article.


\end{document}